\documentclass[twocolumn,pra,superscriptaddress,showpacs]{revtex4}
\usepackage{graphicx}
\usepackage{dcolumn}
\usepackage{amsmath}
\usepackage{bm}
\begin{document}

\title{Autoionization of spin-polarized metastable helium in tight
anisotropic harmonic traps}
\author{Timothy J. Beams}
\affiliation{School of Mathematics, Physics and Information Technology,
James Cook University, Townsville, Australia 4811}
\author{Ian B. Whittingham}
\affiliation{School of Mathematics, Physics and Information Technology,
James Cook University, Townsville, Australia 4811}
\author{Gillian Peach}
\affiliation{Department of Physics and Astronomy, University College London,
Gower Street, London, WC1E 6BT, United Kingdom}

\date{\today}

\begin{abstract}
Spin-dipole mediated interactions between tightly confined metastable
helium atoms couple the spin-polarized quintet ${}^{5}\Sigma^{+}_{g}$ state
to the singlet ${}^{1}\Sigma^{+}_{g}$ state from which autoionization is
highly probable, resulting in finite lifetimes for the trap eigenstates.
We extend our earlier study on spherically symmetric harmonic traps to the
interesting cases of axially symmetric anisotropic harmonic traps and
report results for the  lowest 10 states in "cigar-like" and
"pancake-like" traps with average frequencies
of 100 kHz and 1 MHz. We find that there is a significant suppression
of ionization, and subsequent increase in lifetimes, at trap aspect
ratios $A=p/q$, where $p$ and $q$ are integers, for those states that
are degenerate in the absence of collisions or spin-dipole interactions.
\end{abstract}

\maketitle

\section{Introduction}

There is significant interest in the study and control of quantum
processes involving tightly trapped ultracold neutral atoms where the
trapping environment modifies the collision properties of the
atoms \cite{Nature,Bolda,Peach04}.
Optical lattices, with typical trapping frequencies
of $10^{4}$ to $10^{6}$ Hz, have recently been used to study quantum
phase transitions and other phenomena in alkali systems \cite{Stock}
and there are now proposals to trap metastable helium and neon
in optical lattices \cite{Koel04,Moal06,Vassen06}.
Trapped metastable rare gas atoms offer exciting experimental detection
strategies to
study quantum gases as the large internal energy can be released
through Penning and associative ionization during interatomic collisions
and through ejection of electrons when the atoms strike a metal surface.
The metastable atoms are generally spin-polarized in order to suppress the
autoionization rate and to thereby attain large numbers of trapped atoms.

Spin-dipole interactions between the confined metastable
helium atoms couple the spin-polarized quintet ${}^{5}\Sigma^{+}_{g}$
state to the singlet ${}^{1}\Sigma^{+}_{g}$ state from which Penning
and associative ionization
are highly probable.  This spin-dipole induced coupling results in
finite lifetimes for the trap eigenstates and has been studied for
spherically symmetric harmonic traps by~\cite{Beams04} using a second-order
perturbative treatment of the interaction.

In this paper we extend these calculations to the interesting case of
anisotropic harmonic traps where there arises the possibility of
enhancement and/or suppression of the ionization losses due to
interference between the anisotropic part of the trap potential and the
spin-dipole interaction which are both proportional to a second order
spherical harmonic $Y_{2m}(\theta ,\phi)$. In particular we consider
traps with axial symmetry about the $z$ axis. These traps have $m=0$
and an asymmetry parameter  $\beta $ which ranges from $\beta =2$ for a
"pancake" trap  to $\beta =-1$  for a "cigar" trap.

We have calculated the lowest 10 eigenstates of
$\bar{\omega}/2 \pi = 100$ kHz
and 1.0 MHz traps where $\bar{\omega}$ is the mean angular
frequency of the trap. We find evidence that the
ionization is strongly suppressed and the lifetimes of the trap
states significantly lengthened for particular values of $\beta $,
especially for the higher states considered.

\section{Two-atom collisions in an anisotropic harmonic trap}
\subsection{Hamiltonian}

Consider two trapped atoms with masses $M_{1}$ and $M_{2}$, position vectors
$\bm{r}_{1}$ and $\bm{r}_{2}$, and electronic spin operators
$\bm{S}_1$ and $\bm{S}_2$ subject to a central interatomic potential
$V^{\text{el}}(r)$, where $r=|\bm{r}|=|\bm{r}_{1}-\bm{r}_{2}|$, and a
spin-dipole interaction \cite{Beams04}
\begin{equation}
\label{eq1.1}
H_\text{sd}(\bm{r}) = - \frac{k}{\hbar^2 r^3} \left[
3(\bm{S}_{1}\cdot \hat{\bm{r}})(\bm{S}_{2}\cdot \hat{\bm{r}})
           - \bm{S}_{1} \cdot \bm{S}_{2} \right].
\end{equation}
Here $\hat{\bm{r}} = \bm{r}/r$ is a unit vector
directed along the internuclear axis and
\begin{equation}
\label{eq1.2}
k = \alpha^2 \left( \frac{\mu_e}{\mu_B} \right)^2 E_h a_0^3,
\end{equation}
where $\alpha$ is the fine structure constant, $a_0$ is the Bohr radius,
$(\mu_e/\mu_B)$ is the electron magnetic moment to Bohr magneton ratio
and $E_h = \alpha^{2} m_e c^2 $ is the Hartree energy.
Provided the trapping potential is harmonic, the two-atom Hamiltonian is
separable into center-of-mass and relative motions. The Hamiltonian for
the relative motion is then
\begin{equation}
\label{eq1.3}
H = -\frac{\hbar^{2}}{2M}\;\nabla^{2}_{r}+ V^{\text{el}}(r) +
H_{\text{sd}}(\bm{r}) +V_{\text{trap}}(\bm{r})
\end{equation}
where $M=M_{1}M_{2}/(M_{1}+M_{2})$ is the reduced mass of the system.

The potential for a general anisotropic harmonic trap 
\begin{equation}
\label{eq1.4}
V_{\text{trap}}(\bm{r}) = \frac{M}{2}\;\left(\omega_{x}^{2}x^{2}+
\omega_{y}^{2}y^{2}+\omega_{z}^{2}z^{2}\right),
\end{equation}
where $\omega_{x},\omega_{y}$ and $\omega_{z}$ are the trapping frequencies
in the $x$, $y$ and $z$ directions, can be resolved into isotropic and
anisotropic components by introducing the spherical basis
$x_{\mu }, \mu =0,\pm 1,$ where
\begin{equation}
\label{eq1.5}
x_{\pm } = \mp \frac{1}{\sqrt{2}}(x \pm i y), \quad x_{0} =z.
\end{equation}
Noting that $x_{\mu } = rC^{1}_{\mu}(\theta ,\phi )$ where
\begin{equation}
\label{eq1.6}
   C^l_m (\theta,\phi) \equiv \sqrt{\frac{4\pi}{2l+1}} \;
   Y_{lm} (\theta, \phi)
\end{equation}
is the modified spherical harmonic then, using the product formula
\begin{eqnarray}
\label{eq1.7}
C^{1}_{m_{1}}(\theta ,\phi) C^{1}_{m_{2}}(\theta ,\phi) & = &
\frac{(-1)^{m_{1}}}{3}\;\delta_{m_{1},-m_{2}} \;C^{0}_{0}(\theta , \phi)
\nonumber  \\
&& +\sqrt{\frac{2}{3}}C(1,1,2;m_{1},m_{2},m_{1}+m_{2}) \nonumber  \\
&& \times C^{2}_{m_{1}+m_{2}}(\theta ,\phi)
\end{eqnarray}
where $C(j_{1},j_{2},j;m_{1},m_{2},m)$ is the Clebsch-Gordan coefficient,
$V_{\text{trap}}(\bm{r})$ can be written as
\begin{equation}
\label{eq1.8}
V_{\text{trap}}(\bm{r})  = V_{\text{trap}}^{\text{iso}}(r)
+V_{\text{trap}}^{\text{aniso}}(r,\theta ,\phi ).
\end{equation}
The isotropic component is
\begin{equation}
\label{eq1.9}
V_{\text{trap}}^{\text{iso}}(r) = \frac{M}{2} \bar{\omega}^{2}r^{2}
\end{equation}
and the anisotropic part is
\begin{eqnarray}
\label{eq1.10}
V_{\text{trap}}^{\text{aniso}}(r,\theta ,\phi ) & = &
V_{\text{trap}}^{\text{iso}}(r) \;\left\{\frac{\alpha}{\sqrt{6}}
\left[C^{2}_{2}(\theta ,\phi )
+C^{2}_{-2}(\theta ,\phi )\right]   \right. \nonumber  \\
&&  \left. + \beta \;C^{2}_{0}(\theta, \phi)\right\}.
\end{eqnarray}
We have introduced the mean square frequency
\begin{equation}
\label{eq1.11}
\bar{\omega}^{2} \equiv \frac{1}{3}(\omega_{x}^{2}+\omega_{y}^{2}+
\omega_{z}^{2})
\end{equation}
and the anisotropy parameters
\begin{equation}
\label{eq1.12}
\alpha \equiv 
\frac{\omega_{x}^{2}-\omega_{y}^{2}}{\bar{\omega}^{2}}, \quad
\beta \equiv \frac{1}{3 \bar{\omega}^{2}}(2\omega_{z}^{2}-
\omega_{x}^{2}-\omega_{y}^{2}).
\end{equation}
For axially symmetric traps $\omega_{x}=\omega_{y}$ and $\alpha =0$.
The parameter $\beta $ then varies over the range $-1 \leq \beta \leq 2$ where
$\beta =-1$ corresponds to an infinitely long cigar-shaped trap and
$\beta =2$ to an infinitesimally thin pancake-shaped trap.

\subsection{Perturbation theory}

The spin-dipole interaction is of $O(\alpha^{2})$ and can be treated as a
perturbation.  The unperturbed Hamiltonian $H_{0}$ is then that for
ultracold collisions in an anisotropic harmonic trap.
The two colliding atoms with spin quantum numbers $S_{1}$ and $S_{2}$ are
in the total spin state $|(S_{1}S_{2})SM_{S} \rangle $ where $M_{S}$ is the
projection of $S$ onto the space-fixed $z$ axis.  During the collision they
form the molecular state ${}^{2S+1}\Lambda $ where $\Lambda $ is the
projection of the total electronic orbital angular momentum along the
molecular axis. The adiabatic potential of this state will be denoted by
$V^{\text{el}}_{\Lambda S}(r)$.  The appropriate channel states
are \cite{Beams06}
\begin{equation}
\label{eq2.1}
|\Phi_{a} \rangle = |\Gamma S M_{S} \rangle |lm \rangle
\end{equation}
where $a =\{\Gamma ,S, M_{S},l,m\}$  and $|l m\rangle =Y_{lm}(\theta ,\phi)$
are the relative motion states. The label $\Gamma $
represents the remaining quantum numbers $S_{1},S_{2}, \ldots $ needed to
fully specify the channel.
The unperturbed Hamiltonian $H_{0}$ is given by
\begin{equation}
\label{eq2.2}
H_{0} = - \frac{\hbar^{2}}{2M}\nabla^{2}_{r} + V^{\text{el}}_{\Lambda S}(r)
+V_{\text{trap}}(\bm{r})
\equiv H^{\text{iso}}_{0}+V^{\text{aniso}}_{\text{trap}}(\bm{r})
\end{equation}
and the eigenstates $|\Psi_{n} \rangle $ satisfy
\begin{equation}
\label{eq2.3}
H_{0} | \Psi_{n} \rangle = E_{n} |\Psi_{n} \rangle \,.
\end{equation}
The states $| \Psi_{n}\rangle $ can be expanded in terms of the channel basis defined 
in (\ref{eq2.1}) so that
\begin{equation}
\label{eq2.4}
|\Psi_{n} \rangle = \sum_{a}\;R_{na}(r) |\Phi_{a} \rangle
\end{equation}
and we obtain the matrix eigenvalue equation
\begin{equation}
\label{eq2.5}
\sum_{a} H_{a^{\prime}a} R_{na}(r)= E_{n} R_{na^{\prime}}(r)
\end{equation}
where, assuming the $r$-dependence of the channel states is negligible,
\begin{eqnarray}
\label{eq2.5a}
H_{a^{\prime}a} & = & \langle \tilde{\Phi}_{a^{\prime}}|H_{0}|\Phi_{a}\rangle
\nonumber  \\
& = & h^{\text{iso}}_{a^{\prime}}\delta_{a^{\prime},a} +
\langle \tilde{\Phi}_{a^{\prime}}|V^{\text{aniso}}_{\text{trap}}|\Phi_{a}
\rangle
\end{eqnarray}
and
\begin{eqnarray}
\label{eq2.6}
h^{\text{iso}}_{a} & \equiv & \langle \tilde{\Phi}_{a}|H^{\text{iso}}_{0}|
\Phi_{a} \rangle \nonumber  \\
& = & - \frac{\hbar^2}{2M} \frac{d^2}{dr^2} + \frac{l(l+1)\hbar^2 }{2M r^2} +
V^{\text{el}}_{\Lambda S}(r) \nonumber  \\
&& + \frac{M\bar{\omega}^{2}}{2} r^{2}.
\end{eqnarray}
We have distinguished between right eigenvectors $|\Phi _{a} \rangle $ and
left eigenvectors $\langle \tilde{\Phi}_{a} |$, see \cite{RL},
as $H_{0}$ is non-Hermitian for the singlet state.
In the absence of the anisotropy term equation (\ref{eq2.5}) decouples to
\begin{eqnarray}
\label{eq2.7}
\left[ - \frac{\hbar^2}{2M} \frac{d^2}{dr^2} + \frac{l(l+1)\hbar^2 }{2M r^2} +
V^{\text{el}}_{\Lambda S}(r) + \frac{M \bar{\omega}^{2}}{2} r^{2} \right]
R_{a}(r)   \nonumber  \\
= E_{a} R_{a}(r)\,,
\end{eqnarray}
where $E_{a}=2\hbar \bar{\omega} (n^{*}_{r}+l/2+3/4)$ are the energy
eigenvalues, see \cite{Peach04}, of an isotropic harmonic trap of frequency
$\bar{\omega}$. The quantum number $n^{*}_{r}=n_{r}-\mu$ replaces the
usual integer
$n_{r} \geq 0$ labelling the eigenenergies of a three-dimensional isotropic
oscillator and includes the effect of collisions through the quantum
defect $\mu $ which depends on $\{\Lambda , S, l\}$.

The channel basis elements in (\ref{eq2.5}) can be evaluated using
standard techniques.  From (\ref{eq1.10}) we have
\begin{equation}
\label{eq2.8}
\langle \tilde{\Phi}_{a^{\prime}}| V^{\text{aniso}}_{\text{trap}}
|\Phi_{a} \rangle =
\delta_{\gamma^{\prime},\gamma}\langle l^{\prime}m^{\prime}|
V^{\text{aniso}}_{\text{trap}}|lm \rangle
\end{equation}
where $\gamma =\{\Gamma ,S, M_{S}\}$. Using the Wigner-Eckart theorem
then gives
\begin{equation}
\label{eq2.9}
\langle \tilde{\Phi}_{a^{\prime}}| V^{\text{aniso}}_{\text{trap}}
|\Phi_{a} \rangle =
V^{\text{iso}}_{\text{trap}}(r) D^{\text{trap}}_{a^{\prime}a}
\end{equation}
where
\begin{eqnarray}
\label{eq2.10}
D^{\text{trap}}_{a^{\prime}a} & = & \left\{ \alpha \left[
C(l,2,l^{\prime};m,2,m^{\prime})+C(l,2,l^{\prime};m,-2,m^{\prime})\right]
\right. \nonumber  \\
&& \left. + \beta C(l,2,l^{\prime};m,0,m^{\prime})\right\}
\langle l^{\prime} ||\bm{C}^{2}|| l \rangle
\delta_{\gamma^{\prime},\gamma}.
\end{eqnarray}
The reduced matrix element is given by
\begin{equation}
\label{eq2.11}
\langle l^{\prime}||\bm{C}^2||l \rangle =
\left[ \frac{2l+1}{2l^{\prime}+1} \right]^
  {\frac{1}{2}}\; C(l,2,l^{\prime};0,0,0).
\end{equation}

To second order, the change in energy of the state $|\Psi_{n} \rangle $ due
to the perturbation $H_{\text{sd}}$ is
\begin{equation}
\label{eq2.12}
\Delta E_{n} = \Delta E^{(1)}_{n} + \Delta E^{(2)}_{n},
\end{equation}
where
\begin{equation}
\label{eq2.13}
\Delta E^{(1)}_{n} = \langle \tilde{\Psi}_{n} | H_{\text{sd}}
| \Psi_{n} \rangle
\end{equation}
and
\begin{equation}
\label{eq2.14}
\Delta E^{(2)}_{n} = \sum_{m \neq n}\;\frac{
\langle \tilde{\Psi}_{n}|H^{\dagger}_{\text{sd}}|\Psi_{m} \rangle
\langle \tilde{\Psi}_{m}|H_{\text{sd}}|\Psi_{n} \rangle}
{E_{n}-E_{m}} .
\end{equation}
The evaluation of the first order correction is straightforward. The
second order correction is conveniently evaluated by introducing
\cite{Pert2} the operator $\hat{F_{n}}$ which satisfies
\begin{eqnarray}
\label{eq2.15}
\langle \tilde{\Psi}_{m} |[\hat{F}_{n},H_{0}]|\Psi_{n} \rangle & = &
\langle \tilde{\Psi}_{m} |H_{\text{sd}}|\Psi_{n} \rangle  \\
& = & (E_{n}-E_{m})\langle \tilde{\Psi}_{m} |\hat{F}_{n}|\Psi_{n} \rangle
\end{eqnarray}
for $m \neq n$. Invoking the closure relation
$\sum_{m}|\Psi_{m} \rangle \langle
\tilde{\Psi}_{m}|=I$ then gives
\begin{equation}
\label{eq2.16}
\Delta E^{(2)}_{n} =\langle \tilde{\Psi}_{n} |H_{\text{sd}}\hat{F}_{n}|
\Psi_{n} \rangle -\langle \tilde{\Psi}_{n}|H_{\text{sd}}| \Psi_{n} \rangle
\langle \tilde{\Psi}_{n}|\hat{F}_{n}|\Psi_{n} \rangle .
\end{equation}
Equation (\ref{eq2.15}) yields the inhomogeneous equation
\begin{equation}
\label{eq2.17}
[\hat{F}_{n},H_{0}]|\Psi_{n} \rangle =
(H_{\text{sd}}-\Delta E^{(1)}_{n})|\Psi_{n} \rangle
\end{equation}
so that, expanding $\hat{F}_{n}|\Psi_{n} \rangle $ in the channel basis
(\ref{eq2.1})
\begin{equation}
\label{eq2.18}
\hat{F}_{n}|\Psi_{n} \rangle = \sum_{a}\;f_{na}(r) |\Phi_{a} \rangle
\end{equation}
gives the coupled equations
\begin{eqnarray}
\label{eq2.19}
\sum_{a}[E_{n}\delta_{a^{\prime},a}-H_{a^{\prime}a}]f_{na}(r)  = 
\sum_{a}[\langle \tilde{\Phi}_{a^{\prime}}|H_{\text{sd}}|\Phi_{a} \rangle
\nonumber  \\
 - \Delta E^{(1)}_{n}\delta_{a^{\prime},a}]R_{na}(r).
\end{eqnarray}

To evaluate the elements for the spin-dipole interaction  we note that
the interaction may be written as the scalar product of two
second-rank irreducible tensors, see \cite{Beams04},
\begin{equation}
\label{eq2.20}
H_{\text{sd}} = V_{\text{sd}}(r) \bm{T}^2 \boldsymbol{\cdot} \bm{C}^2,
\end{equation}
where $\bm{T}^2$ is
\begin{equation}
\label{eq2.21}
T^2_{\nu} \equiv \left[\bm{S}^1_1 \boldsymbol{\times}
\bm{S}^1_2 \right]^2_{\nu} = \sum_\mu C(1,1,2;\mu, \nu-\mu, \nu)\;
                         S^1_{1,\mu} S^1_{2,\nu-\mu},
\end{equation}
and $\bm{C}^{2}$ is the second rank tensor formed from the modified
spherical harmonics.
The radial factor is $V_{\text{sd}}(r) = b/r^3$ where
$b \equiv -\sqrt{6}k/\hbar^2$. We then get
\begin{equation}
\label{eq2.22}
\langle \Phi_{a^{\prime}}|H_{\text{sd}}| \Phi_{a}\rangle  =
V_{\text{sd}}(r) D^{\text{sd}}_{a^{\prime}a}
\end{equation}
where, by invoking the Wigner-Eckart theorem,  we find \cite{Beams06}
\begin{eqnarray}
\label{eq2.23}
D^{\text{sd}}_{a^{\prime}a} & = &
\delta_{M_{S^{\prime}} + m^{\prime}, M_S + m}\;
  (-1)^{M_{S^{\prime}} - M_S}  \nonumber  \\*
  && \times  C(S,2,S^{\prime};M_S, M_{S^{\prime}}- M_S,
  M_{S^{\prime}}) \nonumber  \\*
  && \times  C(l,2,l^{\prime}; m, m^{\prime}-m, m^{\prime}) \nonumber \\*
&& \times \langle \Gamma^{\prime} S^{\prime}||\bm{T}^2||\Gamma S \rangle
  \langle l^{\prime}||\bm{C}^2|| l\rangle.
\end{eqnarray}
The reduced matrix element of $\bm{T}^{2}$ is given by
\begin{eqnarray}
\label{eq2.24}
  \langle \Gamma^{\prime} S^{\prime}||\bm{T}^2||\Gamma S \rangle & = &
    \delta_{\Gamma^{\prime},\Gamma}\; \delta_{S_1^{\prime},S_1}\;
    \delta_{S_2^{\prime}S_2}\;\hbar^{2} \nonumber  \\*
&& \times  \sqrt{S_1(S_1+1)S_2(S_2+1)}  \nonumber \\*
&& \times   \left[
                         \begin{array}{lll}
                                 S_1 & S_2 & S \\
                                 1 & 1 & 2 \\
                                 S_1 & S_2 & S^{\prime}
                         \end{array}
\right],
\end{eqnarray}
where the angular momentum coefficient in (\ref{eq2.24}) is related to the
Wigner $9-j$ coefficient
\begin{eqnarray}
\label{eq2.25}
\left[  \begin{array}{lll}
                         j_1 & j_2 & j \\
                         k_1 & k_2 & k \\
                         j_1^{\prime} & j_2^{\prime} & j^{\prime}
     \end{array} \right]
  & \equiv &
[(2j_1^{\prime} + 1)(2j_2^{\prime} + 1)(2j + 1)(2k + 1)]^{\frac{1}{2}}
\nonumber  \\*
&& \times   \left\{
                         \begin{array}{lll}
                                 j_1 & j_2 & j \\
                                 k_1 & k_2 & k \\
                                 j_1^{\prime} & j_2^{\prime} & j^{\prime}
                         \end{array}
                 \right\}.
\end{eqnarray}

In terms of the channel basis expansions, the energy shifts are given by
\begin{equation}
\label{eq2.26}
\Delta E^{(1)}_{n} = \sum_{a^{\prime},a}\;D^{\text{sd}}_{a^{\prime}a}
\langle \tilde{R}_{na^{\prime}}|V_{\text{sd}}|R_{na} \rangle
\end{equation}
and
\begin{equation}
\label{eq2.27}
\Delta E^{(2)}_{n} = \sum_{a^{\prime},a}\;D^{\text{sd}}_{a^{\prime}a}
\langle \tilde{R}_{na^{\prime}}|V_{\text{sd}}|f_{na} \rangle
-\Delta E^{(1)}_{n}\sum_{a}\langle \tilde{R}_{na}|f_{na}\rangle .
\end{equation}

\section{Application to ultracold spin-polarized metastable helium atoms}

We now consider application of the theory to spin-polarized
metastable helium atoms tightly confined in axially symmetric harmonic
traps for which the asymmetry parameter $\alpha $ in (\ref{eq1.12})
vanishes. The colliding atoms are in the
$|(S_{1}S_{2})SM_{S}\rangle =|(11)22 \rangle $ ${}^5\Sigma_g^+$
molecular state but, noting that the reduced matrix elements
for the spin-dipole interaction are  \cite{Beams06}
$\langle \Gamma^{\prime} 2||\bm{T}^2||\Gamma 2\rangle=
\delta_{\Gamma^{\prime},\Gamma} \;
\hbar^2\sqrt{{7}/{3}}$ and
$\langle \Gamma^{\prime} 0||\bm{T}^2||\Gamma 2\rangle= -
\delta_{\Gamma^{\prime},\Gamma}\; \hbar^2 \sqrt{{5}/{3}}$
($\langle \Gamma^{\prime} 1||\bm{T}^2||\Gamma 2\rangle$ vanishes by
symmetry of the $9-j$ symbol), then the spin-dipole interaction couples
this state to the
$|(11)00 \rangle $ ${}^{1}\Sigma^{+}_{g}$ state from which
Penning and associative ionization
processes are highly probable. We model these ionization processes
from the singlet state using a complex optical potential of the form
$V(r)-i\Gamma(r)/2$ where $\Gamma(r)$ is the ionization width.
This coupling between the initial quintet state and the singlet state
produces complex energy shifts
$\Delta E_{n} = \Delta E_{n}^{\text{re}} - i\gamma_{n}/2$
due to the complex form of the singlet potential.
The $1/e$ lifetimes of the trap states will be
$\tau_{n} =\hbar/\gamma_{n}$ \footnote{Note that the lifetimes given
in \cite{Beams04} for a spherically symmetric trap
correspond to $1/e^{2}$ decay}.

The eigenvalue equations (\ref{eq2.5}) for $R_{na}(r)$
and the coupled differential equations (\ref{eq2.19}) for the perturbed
functions $f_{na}(r)$ were solved using a discrete
variable representation and a scaled radial coordinate grid
$\rho =r/\xi = \zeta t^{\sigma}$ where $\xi = \sqrt{\hbar/M\bar{\omega}}$ is
the effective range of the oscillator ground state and $\zeta $ and $\sigma $
are scaling parameters (see \cite{Peach04} for details).
We note that the bosonic symmetry of the identical metastable
helium atoms restricts the scattering to even partial waves in the
expansions (\ref{eq2.4}) and (\ref{eq2.18}).

As input to the problem we require the Born-Oppenheimer potentials for
both the ${}^5\Sigma_g^+$ and ${}^1\Sigma_g^+$ electronic states of the
metastable helium dimer.  For the ${}^5\Sigma_g^+$ potential we have
used the full  potential $V^{5}_{\text{PJ}}(r)$
of Przybytek and Jeziorski~\cite{Przyb05} which includes adiabatic and
relativistic corrections, adjusted to match the
experimental binding energy of the least bound ($v=14$)
state~\cite{Moal06}.  For the singlet potential we use a potential
$V^{1}_{\text{M}}$ constructed from the short-range potential of
M\"uller \textit{et al}~\cite{Muller91} exponentially
damped onto the quintet potential at long range \cite{Venturi99}.
Also required is the ionization width $\Gamma(r)$. Two forms
were used, $\Gamma_{\text{M}}(r)$  obtained from a least squares fit to the
tabulated results in~\cite{Muller91} and the simpler form
$\Gamma_{\text{GMS}}(r)=0.3 \exp(-r/1.086)$ advocated
in~\cite{Garrison73}.

\section{Results and Discussion}

We have calculated the lifetimes of the lowest 10 trap states for
axially symmetric anisotropic harmonic traps with average trapping
frequencies of 100 kHz and 1 MHz. We label these states $v=0, \ldots , 9$.

With the power law scaling $\sigma=10$, three to four digit convergence with
respect to the radial grid was obtained using 500 (450) grid points for
$\bar{\omega}/2 \pi = 100$ kHz (1 MHz) and an outer boundary set at
$15 \xi$. The position of the outer boundary was determined by the
requirement  that the states up to at least $v=9$ were insensitive to
its position.  A similar three to four digit convergence with respect to
the number of partial waves included was generally obtained with six partial
waves ($l=0,2, \ldots , 10$) for asymmetry values in the range
$-0.9 \leq \beta \leq 1.9$, where $\Delta \beta =0.01$.
However, for some of the  cases where the lifetimes are strongly enhanced
(see below), satisfactory partial wave convergence was not obtained and
the results reported for these cases should be regarded as indicative only.

\begin{figure*}
\includegraphics[width=0.7\linewidth]{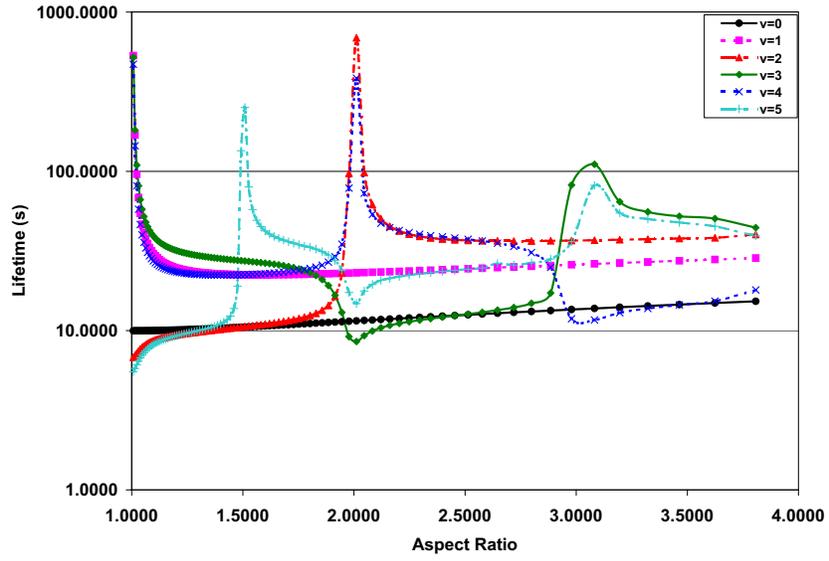}
\caption{\label{fig1}Lifetimes of the six lowest trap states for
"cigar-like" traps with $\bar{\omega}/2 \pi =100$ kHz
as a function of trap aspect ratio $A=\omega_{x}/\omega_{z}$.}
\end{figure*}
\begin{figure*}
\includegraphics[width=0.7\linewidth]{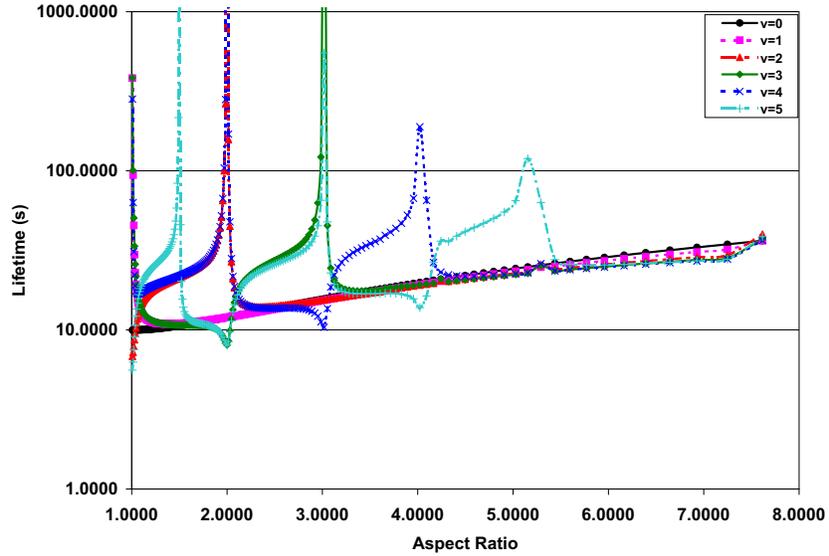}
\caption{\label{fig2}Lifetimes of the six lowest trap states for
"pancake-like" traps with $\bar{\omega}/2 \pi =100$ kHz
as a function of trap aspect ratio $A=\omega_{z}/\omega_{x}$.}
\end{figure*}

The lifetimes  of the six lowest states for $\bar{\omega}/2\pi =100$ kHz
are shown in Figs \ref{fig1} and \ref{fig2} as a function of
the trap aspect ratio $A$. For "cigar-like" traps with $\omega_{x} >
\omega_{z}$ the aspect ratio is
\begin{equation}
\label{eq3.2a}
A = \frac{\omega_{x}}{\omega_{z}} = \sqrt{\frac{(1-\beta /2)}{(1+\beta )}}
\end{equation}
whereas for "pancake-like" traps with $\omega_{z} > \omega_{x}$ it is
$A=\omega_{z}/\omega_{x}$.


The results display a strong sensitivity of most lifetimes to the aspect
ratio $A$, especially around the values $A=p/q$
where $p$ and $q$ are integers.
Under higher resolution these peaks show a double peak structure
centred on $A=p/q$.
In Table 1 results are presented for the
10 lowest states at the $\beta $ values  closest to these aspect ratios and
it is clear that there is an aspect ratio for nearly every state at which
the lifetime is greatly enhanced.
The behavior near $A=1$ ($\beta =0$) is also interesting. For $\beta =0$ the
states become pure partial waves and those states $v=0,2,5,9$ which
contain $l=0$ develop lifetimes which increase slowly for small
$|\beta |$ . However the other states show strongly enhanced
lifetimes at very small values of $\beta \sim O(10^{-3})$
either side of $\beta =0$.  No results are reported for the $v=4,7,8$
states at $A=1$ as for these cases our program developed numerical
instabilities due to ill-conditioned matrices at very small
$\beta \leq O(10^{-4})$.
We note that the
results given in \cite{Beams04} for the $l=0$ states of a spherically
symmetric trap are reproduced for $\beta =0$ if we use the
potential $V^{5}_{\text{SM}}(r)$ of St\"arck and Meyer~\cite{Starck94}.

\begin{table*}
\caption{\label{lifetimes} Lifetimes (in seconds)
of the 10 lowest trap states for axially symmetric  anisotropic
harmonic traps of selected aspect ratios $A$ and an
average trapping  frequency of 100 kHz.
The upper part of the table is for "cigar-like" traps  where
$A=\omega_{x}/\omega_{z}$, the lower part is for "pancake-like" traps
where $A=\omega_{z}/\omega_{x}$. The numbers in parentheses denote powers of
10. For many of the cases of very strongly enhanced lifetimes the results
are indicative only as satisfactory partial wave convergence was not
achieved.}
\begin{ruledtabular}
\begin{tabular}{lcccccccccc}
$A$ & $v=0$ &  $v=1$ & $v=2$ & $v=3$ & $v=4$ & $v=5$ & $v=6$
& $v=7$ & $v=8$ & $v=9$ \\
\hline
1     & 9.977 & 5.76(-6) & 6.714 & 1.35(-5) &  & 5.399 & 2.28(-5) &  &
& 4.645 \\
1.25  &  10.14 &  23.86  & 9.604   & 30.55  & 22.83  & 9.618   & 34.19  & 29.36  & 22.80  &  10.04 \\
1.5   &  10.50 &  22.44  & 10.54  & 27.49  & 22.42  & 134.8 & 8.592   & 27.44  & 93.43  &  17.31 \\
1.75  &  10.98 &  22.53  & 11.91  & 24.39  & 23.90  & 34.66  & 11.94  & 26.20  & 33.67 &  24.99  \\
2     &  11.51 &  23.04  & 689.5  & 8.540  &  384.8 & 14.78  & 1.144(3) & 941.1 & 7.135  &  554.3 \\
2.25  &  12.01 &  23.67  & 40.32  & 11.38  & 41.24  & 22.32  & 53.39  & 36.95 &  11.28 &  55.80  \\
2.5   &  12.57 &  24.46  & 36.83  & 12.70  & 37.42  & 24.29  & 47.06  & 34.21 &  1.76(3) & 10.32 \\
2.75  &  13.01 &  25.09  & 36.50  & 13.96  & 33.76  & 25.80  & 44.35  & 33.79 &  63.40 &  13.22  \\
3     &  13.56 &  25.91  & 36.76  & 82.08  & 11.86  & 36.17  & 30.92  & 36.43 &  49.55 &  16.60  \\
3.25  &  14.02 &  26.60  & 37.20  & 64.36  & 12.95  & 54.67  & 24.51  & 58.80 &  31.72 &  95.24  \\
3.5   &  14.57 &  27.44  & 37.86  & 52.19  & 14.49  & 47.73  & 26.70  & 55.57 &  31.84 &  73.23  \\
3.75  &  15.27 &  28.56  & 40.11  & 44.31  & 18.01  & 39.35  & 24.84  & 59.83 &  27.37 &  66.31  \\
        
&&&&&&&&&&\\                                                                         
										
1.25  &  10.16 &  11.08  & 18.00  & 11.24  & 19.14  & 25.52  & 11.14  & 19.36 &  26.77 &  35.07 \\
1.5   &  10.60 &  10.95  & 21.28  & 10.73  & 21.73  & 2.2(2) & 8.133   & 21.34 &  206.9 & 8.306  \\
1.75  &  11.22 &  11.36  & 25.85  & 10.62  & 26.04  & 10.90  & 32.11  & 25.00 &  10.44 &  34.55  \\
2     &  11.98 &  11.99  & 4.0(6) & 8.042  & 6.1(6) & 8.017  & 4.2(3) & 3.7(2) & 6.451 &  2.2(2)  \\
2.25  &  12.80 &  12.71  & 14.08  & 22.30  & 13.94  & 21.50  & 14.89  & 21.28 &  29.26 &  14.34  \\
2.5   &  13.67 &  13.50  & 14.12  & 27.25  & 13.67  & 25.73  & 14.09  & 23.91 &  96.42 &  10.96  \\
2.75  &  14.60 &  14.36  & 14.71  & 33.25  & 13.67  & 29.99  & 13.95  & 28.57 &  13.88 &  34.22  \\
3     &  15.55 &  15.23  & 15.41  & 1.2(2) & 11.23  & 65.20  & 11.73  & 39.89 &  12.46 &  52.50  \\
3.25  &  16.57 &  16.19  & 16.21  & 17.98  & 28.07  & 17.24  & 26.14  & 17.64 &  22.06 &  18.29  \\
3.5   &  17.67 &  17.19  & 17.06  & 17.65  & 34.87  & 16.86  & 30.27  & 17.24 &  25.13 &  17.06  \\
3.75  &  18.79 &  18.28  & 18.21  & 18.53  & 43.46  & 16.95  & 39.90  & 16.94 &  33.72 &  17.01  \\
4     &  19.87 &  19.29  & 19.12  & 19.23  & 1.9(2) & 13.82  & 82.23  & 14.63 &  34.53 &  17.59  \\
4.25  &  20.82 &  20.23  & 20.42  & 20.95  & 23.07  & 36.13  & 22.20  & 26.31 &  20.65 &  23.55  \\
4.5   &  21.94 &  21.22  & 20.82  & 20.65  & 21.71  & 41.38  & 20.64  & 38.00 &  20.14 &  31.79  \\
4.75  &  23.27 &  22.45  & 21.88  & 21.56  & 22.03  & 50.10  & 20.45  & 43.68 &  20.16 &  35.71  \\
5     &  24.30 &  23.40  & 22.68  & 22.26  & 22.50  & 64.85  & 19.68  & 51.71 &  19.62 &  2.2(3) \\

\end{tabular}
\end{ruledtabular}
\end{table*}

To understand this behavior  we note
that, in the absence of collisions, the energy eigenvalues for an
axially symmetric harmonic trap
\begin{equation}
\label{eq3.1}
E_{0}^{\text{no-coll}} = \hbar \omega_{x}(n_{x}+\frac{1}{2}) +
\hbar \omega_{y}(n_{y}+\frac{1}{2})
+\hbar \omega_{z} (n_{z}+\frac{1}{2})
\end{equation}
can be written as
\begin{equation}
\label{eq3.2}
E_{0}^{\text{no-coll}} = \frac{\hbar \bar{\omega}}{2}
\sqrt{1+\beta }[2n_{z}+1 +2A(n_{x}+n_{y}+1)]
\end{equation}
where $A = \omega_{x}/\omega_{z}$. For $A=p/q$ this becomes
\begin{equation}
\label{eq3.3}
E_{0}^{\text{no-coll}} =\frac{\hbar \bar{\omega}}{2}
\sqrt{\frac{3}{q^{2}+2p^{2}}} [(2n_{z}+1)q +2(n_{x}+n_{y}+1)p].
\end{equation}
From (\ref{eq3.3}) it is clear that, for a given pair of integers $(p,q)$,
there will be states $(n_{x},n_{y},n_{z})$ that are degenerate. For
example, consider the case $(p,q)=(3,2)$ for which the lowest
degenerate states
$(n_{x},n_{y},n_{z})$ are $(1,1,0),(0,0,3)$, corresponding to
the vibrational states $v=(5,6)$ and $(1,1,1),(0,0,4)$ corresponding to
$v=(8,9)$. Collisions then break the  degeneracies of the eigenstates formed
from even partial waves.
Inspection of Figs \ref{fig1} and \ref{fig2} and Table 1 shows that
it is these degenerate oscillator states that have the strongly
enhanced lifetimes.\\

The enhancement of some lifetimes at very small $\beta $ can also be traced
to underlying degeneracies.
As $\beta  \rightarrow 0$, the energies approach the isotropic oscillator
eigenvalues
\begin{equation}
\label{eq3.4}
E_{0}^{\text{no-coll}}(\beta =0) =\frac{\hbar \bar{\omega}}{2}(4n_{r}+2l+3)
\end{equation}
which have the degeneracies $(n_{r},l)= (0,2), (1,0)$ corresponding
to $v=1,2$, $(0,4),(1,2),(2,0)$ to $v=3,4,5$ and $(0,6),(1,4),(2,2),(3,0)$
to $v=6, \ldots ,9$. Again collisions break some of these
degeneracies, predominantly by raising the energies of the $l=0$
states $v=0,2,5,9$.

The results for $\bar{\omega}/2\pi =1$ MHz show the same general features
as those for 100 kHz and are not presented here.

In conclusion, our calculations show that the ionization losses are
sensitive to the anisotropic trapping environment and that
significant suppression of ionization can occur at particular trap aspect
ratios. Although there is some dependance of our results on the choice
of ionization width $\Gamma (r)$ (the differences in lifetimes obtained using
the ionization widths
$\Gamma_{\text{M}}$ and $\Gamma_{\text{GMS}}$ are remarkably constant
with the $\Gamma_{\text{GMS}}$ results lower by 14.2\%) and 
satisfactory partial wave convergence could not obtained with the limit of
six partial waves  for some of the
strongly enhanced lifetimes, the general features displayed by our results
are robust.

\end{document}